\documentclass[aps,prx,twocolumn,superscriptaddress]{revtex4-2}
\usepackage[normalem]{ulem}
\usepackage{amsmath}
\usepackage{amsfonts}
\usepackage{amssymb}
\usepackage{graphicx}
\usepackage[colorlinks=True,citecolor=myblue,linkcolor=myorange,urlcolor=mygreen]{hyperref}
\usepackage{hypcap}
\usepackage{braket}
\usepackage{bm}
\usepackage{bbm}
\usepackage{tikz}
\usepackage[export]{adjustbox}
\usepackage{xcolor}
\pagecolor{white}
\usepackage{mathtools}
\mathtoolsset{centercolon}

\definecolor{myblue}{RGB}{0,0,130}
\definecolor{myorange}{RGB}{130,50,0}
\definecolor{mygreen}{RGB}{0,130,0}

\usepackage{tikz}
\usetikzlibrary{arrows,decorations.pathmorphing}

\begin{document}

\title{Probing post-measurement entanglement without post-selection}
\author{Samuel J. Garratt}
\affiliation{Department of Physics, University of California, Berkeley, California 94720, USA}
\author{Ehud Altman}
\affiliation{Department of Physics, University of California, Berkeley, California 94720, USA}
\affiliation{Materials Science Division, Lawrence Berkeley National Laboratory, Berkeley, California 94720, USA}

\begin{abstract}
We study the problem of observing quantum collective phenomena emerging from large numbers of measurements. These phenomena are difficult to observe in conventional experiments because, in order to distinguish the effects of measurement from dephasing, it is necessary to post-select on sets of measurement outcomes whose Born probabilities are exponentially small in the number of measurements performed. An unconventional approach, which avoids this exponential `post-selection problem', is to construct cross-correlations between experimental data and the results of simulations on classical computers. However, these cross-correlations generally have no definite relation to physical quantities. We first show how to incorporate shadow tomography into this framework, thereby allowing for the construction of quantum information-theoretic cross-correlations. We then identify cross-correlations which both upper and lower bound the measurement-averaged von Neumann entanglement entropy, and additional cross-correlations which lower bound the measurement-averaged purity and entanglement negativity. These bounds show that experiments can be performed to constrain post-measurement entanglement without the need for post-selection. To illustrate our technique we consider how it could be used to observe the measurement-induced entanglement transition in Haar-random quantum circuits. We use exact numerical calculations as proxies for quantum simulations and, to highlight the fundamental limitations of classical memory, we construct cross-correlations with tensor-network calculations at finite bond dimension. Our results reveal a signature of measurement-induced criticality that can be observed using a quantum simulator in polynomial time and with polynomial classical memory. \end{abstract}

\maketitle

\section{Introduction}

Local measurements allow for the protection and control of quantum correlations, and are essential for error-corrected quantum computation \cite{nielsen2000quantum}. This aim has motivated the development of local readout techniques in quantum simulation platforms from trapped ion \cite{monroe2021programmable} and neutral atom arrays \cite{browaeys2020many} to superconducting quantum processors \cite{kjaergaard2020superconducting}. New questions also arise in many-body quantum mechanics, in particular about the effects of many measurements on entangled quantum states. Examples of phenomena arising in this setting are the measurement-induced entanglement transitions in quantum circuits \cite{li2018quantum,skinner2019measurement,potter2022entanglement,fisher2023random}, as well as the creation \cite{raussendorf2005longrange,piroli2021quantum,tantivasadakarn2022longrange,lu2022measurement,zhu2022nishimoris,lee2022decoding,tantivasadakarn2022hierarchy,fossfeig2023experimental,iqbal2023topological} and restructuring \cite{garratt2022measurements,lee2023quantum,weinstein2023nonlocality,yang2023entanglement,murciano2023measurementaltered,sun2023new} of entangled states through measurement.

\begin{figure}
	\includegraphics[width=0.47\textwidth]{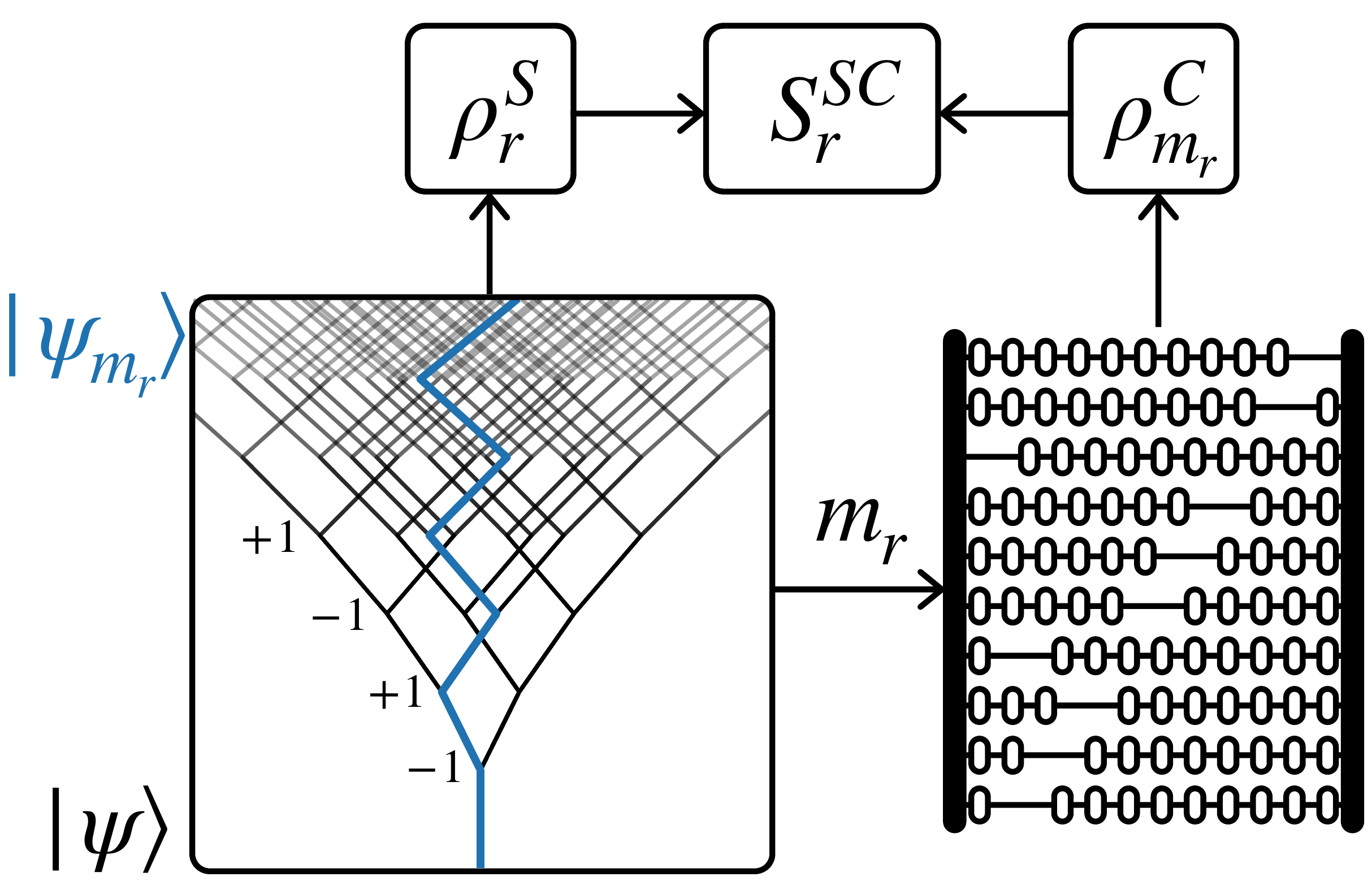}
\caption{Illustration of the protocol for a single run $r$. The branching process on the left represents an experiment involving measurements that takes the system from initial state $\ket{\psi}$ to post-measurement state $\ket{\psi_{m_r}}$, where $m_r$ is the set of $M$ measurement outcomes observed in run $r$ (here indicated by $\pm 1$). In each run one extracts a single shadow $\rho^S_r$ of a subsystem density matrix $\rho_{m_r}$. The outcomes $m_r$ are used as input to a calculation on a classical computer (represented by the abacus), and this calculation returns an estimate $\rho^C_{m_r}$ for the density matrix $\rho_{m_r}$. Finally one constructs the object $S^{S C}_r = -\text{Tr}[\rho^S_r \log \rho^C_{m_r}]$. Averaging $S^{S C}_r$ over runs of the experiment generates an upper bound on the measurement-averaged entanglement entropy of the density matrices $\rho_{m_r}$.}
\label{fig:abacus}
\end{figure}

Experiments studying the effects of many measurements must, however, be conducted in a fundamentally different way. This is because, to characterize a quantum state in a conventional experiment, we must prepare and measure it more than once. If the preparation of this state involves measurements, we are then forced to post-select for sets of outcomes whose probabilities are exponentially small in the number of measurements performed, and hence the time required is exponentially large \cite{koh2022experimental}. Moreover, averaging over runs of an experiment has the effect of converting measurements into dephasing events, obscuring structures in the ensemble of post-measurement states.

Two different strategies have emerged for avoiding this `post-selection problem', each involving a simulation on a classical computer (a `classical simulation'). One possibility is to use experimental measurement outcomes as input to the calculation of a unitary operation which decodes quantum information \cite{gullans2020scalable,noel2022measurement,dehghani2022neuralnetwork}; this protocol is analogous to the decoding step in quantum teleportation~\cite{bennett1993teleporting}. The second strategy, which is the focus of this work, is to calculate cross-correlations between the results of experiments and classical simulations of the system \cite{li2021robust,garratt2022measurements,lee2022decoding,hoke2023quantum}. This strategy has the advantage that the effects of measurements can be inferred in classical post-processing, and hence the duration of the classical simulation is not limited by experimental decoherence time scales. The idea is closely related to cross-entropy benchmarking, studied in the contexts of random circuit sampling \cite{arute2019quantum,arute2019quantum} and the measurement-induced entanglement transition \cite{li2022cross}, with the important difference that the cross-correlations of Refs.~\cite{li2021robust,garratt2022measurements,lee2022decoding,hoke2023quantum} are designed to probe properties of post-measurement quantum states rather than statistical properties of the measurement record.  

Ultimately, however, the aim of experiment should be to extract information intrinsic to the system. The problem with constructing probes that depend on classical simulations, based on either decoding or cross-correlation, is that the results depend sensitively on how we choose to model the system. Given the apparent necessity of involving a classical simulation, it is natural to ask whether it is even possible to probe intrinsic properties of post-measurement quantum states without post-selection.  

In this work we show that cross-correlations between experimental data and classical simulations can be used to bound model-independent properties of post-measurement states. Our cross-correlations are inspired by shadow tomography \cite{huang2020predicting,elben2023randomized}; shadows of a density matrix have the property that, when suitably averaged, they reproduce the density matrix, but in the regime where there is a post-selection problem we will not be able to perform this average. Nonetheless, we will show that cross-correlations taking shadows of post-measurement states as input can be used to construct an estimate for the measurement-averaged entanglement entropy [see Fig.~\ref{fig:abacus}]. By identifying this estimate as a contribution to an (averaged) quantum relative entropy, which is non-negative, we arrive at a resource-efficient upper bound on the true measurement-averaged entanglement entropy. This bound becomes an equality for a perfect classical simulation. Using the monotonicity of the quantum relative entropy, we also construct an upper bound on the conditional entropy \cite{nielsen2000quantum}. Recasting this as a lower bound on the entanglement entropy, we identify physical settings where the entanglement can be determined to within a threshold set by the accuracy of the classical simulation. 

While a quantum simulation can represent a highly-entangled many-body state, the classical memory required for this is in general exponentially large in the entanglement entropy. This difference between quantum and classical memory imposes a fundamental obstacle to schemes involving classical simulations. To investigate the effects of a constrained classical memory we consider an application of our approach in the context of the measurement-induced entanglement transition \cite{li2018quantum,skinner2019measurement,potter2022entanglement,fisher2023random}. In place of a quantum simulation we use exact numerics to study the dynamics of one-dimensional chains of $L \leq 24$ qubits, and we cross-correlate the results with matrix-product state (MPS) simulations at finite bond dimension $\chi$. The MPS representation only occupies classical memory $\sim L \chi^2$, to be contrasted with $\sim 2^L$ for a generic many-body quantum state. Increasing $\chi$ improves the simulation, causing our upper bound on the measurement-averaged entanglement to become more restrictive. In the vicinity of measurement-induced criticality we show numerically that the separation between the upper bound and the exact value decays exponentially $\sim e^{-\chi/\chi^{QC}(L)}$ with increasing $\chi$, and that the decay constant $\chi^{QC}(L)$ grows as a power of $L$. This result suggests that the generic measurement-induced entanglement transition can be observed in experiment with computational resources that scale only polynomially with system size. 

The fact that the measurement-averaged entanglement entropy can be bounded without post-selection has more general implications. For example, it opens the door to experiments on deep thermalization \cite{cotler2023emergent,choi2023preparing,ho2022exact,ippoliti2023dynamical,claeys2022emergent,mcginley2022shadow,lucas2023generalized}, measurement-induced teleportation \cite{bao2022finite,lin2022probing,antonini2022holographic,antonini2023holographic}, and measurement-altered criticality \cite{garratt2022measurements,lee2023quantum,weinstein2023nonlocality,yang2023entanglement,murciano2023measurementaltered,sun2023new}. Moreover, by converting the post-selection problem into a variational problem, our work suggests a way to improve classical simulations using experimental data.

It is important to separate our approach from adaptive schemes. If individual measurement outcomes are used as input to local unitary feedback operations, nontrivial correlations \cite{piroli2021quantum,lu2022measurement,tantivasadakarn2022hierarchy,fossfeig2023experimental,iqbal2023topological,lu2023mixedstate} and dynamics \cite{sierant2022dissipative,mcginley2022absolutely,friedman2022measurementinduced,iadecola2022dynamical,buchhold2022revealing} may be probed without post-selection, i.e. through the density matrix averaged over runs of the experiment. With or without feedback, such an average washes out the features in the ensemble of post-measurement states that are of interest here (such as measurement-induced entanglement transitions \cite{odea2022entanglement,ravindranath2022entanglement,piroli2022triviality,sierant2023controlling}). A key physical difference is that averaging over the outcome of a measurement converts it into a quantum channel and, although local measurements can have nonlocal effects on entangled states, local quantum channels cannot.

This paper is organized as follows. In Sec.~\ref{sec:postselection} we outline the origin of the post-selection problem, and in Sec.~\ref{sec:scalar} we describe quantum-classical cross-correlations. There we also provide a first example of how these cross-correlations can be used to bound physical quantities. In Sec.~\ref{sec:shadows} we discuss shadows, use them to construct matrix-valued cross-correlations, and then derive a bound on the measurement-averaged purity. Following this, in Sec.~\ref{sec:entanglement}, we introduce resource-efficient bounds on the measurement-averaged von Neumann entanglement entropy, and in Sec.~\ref{sec:variance} we discuss their statistical fluctuations. Section~\ref{sec:transition} addresses the measurement-induced entanglement transition, and in Sec.~\ref{sec:decoding} we discuss the connection between our work and Ref.~\cite{gullans2020scalable}. In Sec.~\ref{sec:negativity} we then present a bound on the measurement-averaged entanglement negativity. We summarize our results, and offer an outlook, in Sec.~\ref{sec:discussion}.

\section{Post-selection problem}\label{sec:postselection}
Here we discuss the post-selection problem. We consider experiments involving $M \gg 1$ measurements, having sets of outcomes $m$, which create post-measurement quantum states that depend on $m$. For example, if we have a system of $L$ qubits in a random pure state, and we measure the Pauli $Z$ operator on $M=L-2$ of the qubits, finding outcomes $m$, we are left with a random pure state on the two unmeasured qubits. We could then ask how to characterize the post-measurement density matrix $\rho_m$ of one of these two qubits. More generally, the post-measurement density matrices $\rho_m$ discussed in this work will be taken to describe states of a number of qubits of order unity. 

To characterize the density matrix $\rho_{m}$ we must measure an observable, and this returns one of the eigenvalues of the observable. To determine the post-measurement expectation value, on the other hand, we must prepare $\rho_{m}$ multiple times. The problem is that the probability $P_m$ to observe $m$ is in general exponentially small in $M$. If we repeat the experiment a finite number of times $R$, then for large $M$ we have typically have $P_m R \ll 1$, and hence we expect to find each $m$ no more than once. Therefore, we cannot characterize the ensemble of $\rho_m$ in a conventional way. Throughout this work, we will label the runs of the experiment $r=0,\ldots,(R-1)$, and the corresponding outcomes $m_r=(m_{r,0},\ldots,m_{r,M-1})$. `Monte Carlo' averages over $R \gg 1$ runs of the experiment will be denoted by $\mathbbm{E}_r[\cdots]$, while averages over the ensemble of all outcomes (weighted by their respective Born probabilities $P_m$) will be denoted $\mathbbm{E}_m[\cdots]$.

Suppose that in each run $r$ of the experiment we probe a post-measurement density matrix $\rho_{m_r}$ by measuring a Pauli matrix $Z$. This `probe' measurement should be distinguished from the $M$ `preparation' measurements which generate the ensemble of states that we hope to characterize. The result of the probe measurement is $z_r = \pm 1$, corresponding to state $\ket{z_r}$, but if we average $z_r$ over runs we find only a property of the `dephased' density matrix $\mathbbm{E}_m[\rho_m]=\sum_m P_m \rho_m$,
\begin{align}
    \mathbbm{E}_r[z_r] = \mathbbm{E}_m[\langle Z \rangle_m] = \text{Tr}\big[\mathbbm{E}_m[\rho_m]Z\big],
    \label{eq:Z1}
\end{align}
where $\langle Z \rangle_m = \text{Tr}[\rho_m Z]$. Here the average over runs, which occurs in classical post-processing, has the effect of converting our measurements into dephasing. This is because we chose to average an object that is linear in the post-measurement density matrices. Averages which distinguish the effects of measurement from dephasing are nonlinear in the post-measurement density matrices, but such averages cannot be determined without post-selection. For example, there is no way to construct an average which converges to $\mathbbm{E}_m[\braket{Z}^2_m]$ if we have only one eigenvalue $z_r$ for each observed outcome $m_r$.

Here we have described the origin of the post-selection problem, which arises for large $M$. The effects of measurements are physically distinct from dephasing, but standard averages over results of our experiment are blind to this difference. Our limitation is that we have access only to sets of outcomes $m_r$ and the results of our `probe' measurements, such as $z_r$ above. In the next section we describe how to use this information to detect structure in the ensemble of post-measurement density matrices.

\section{Cross-correlations}\label{sec:scalar}

To understand quantum-classical cross-correlations, let us return to the above example of preparing post-measurement density matrices $\rho_{m_r}$ and trying to characterize them by measuring $Z$. Our discussion in this section largely follows Ref.~\cite{garratt2022measurements}. It will be convenient to write the observed eigenvalue of $Z$ as 
\begin{align}
    z_r = \braket{Z}_{m_r}+[z_r -\braket{Z}_{m_r}].\label{eq:zdecompose}
\end{align} 
Equation~\eqref{eq:Z1} simply tells us that averaging over runs of the experiment washes out the random fluctuations ${[z_r -\braket{Z}_{m_r}]}$ around the mean $\braket{Z}_{m_r}$. The problem is that, through this average, we have also lost information on the variations of $\braket{Z}_{m_r}$ across different runs. By cross-correlating the results $z_r$ of our experiment with numbers $w_{m_r}$ that depend on the outcomes $m_r$ but not on $z_r$, we can preserve this information,
\begin{align}
    \mathbbm{E}_r[w_{m_r} z_r] = \mathbbm{E}_m[w_m \langle Z \rangle_m]. \label{eq:wm}
\end{align}
There is a lot of freedom in how the weights $w_m$ are chosen. One basic requirement is that $w_m$ depends on $m$, since otherwise Eq.~\eqref{eq:wm} reduces to Eq.~\eqref{eq:Z1}. Another is that the variance of $w_{m_r} z_r$ should grow no faster than polynomially with $M$, and this requirement rules out $w_{m_r}=\delta_{m_r m}$, which corresponds to post-selecting for outcomes $m$.

With a suitable choice of $w_m$, we should be able to learn something about the ensemble of $\rho_{m}$. In Ref.~\cite{garratt2022measurements} we advocated for $w_m$ equal to an estimate for the post-measurement expectation value of an observable, for example $w_m = \braket{Z}_m^C$. In principle, this estimate is determined via a calculation on a classical computer which takes the experimental measurement outcomes $m$ as input. The operator used to construct $w_m$ need not be the same as the operator one measures in experiment: the authors of Refs.~\cite{li2021robust,hoke2023quantum} chose $w_m$ to be equal to the sign of the expectation value $\braket{Z}_m^C$.

By choosing such a $w_m$, the experimentally-accessible probe of the measured quantum system becomes a cross-correlation between experiment and theory. With $w_m = \braket{Z}_m^C$ we have
\begin{align}
    \mathbbm{E}_r[ \braket{Z}_{m_r}^C z_r] = \mathbbm{E}_m[  \braket{Z}_{m}^C \braket{Z}_m]  \label{eq:rmaverage}
\end{align}
This quantum-classical cross-correlation can then be compared with the result of the simulation 
\begin{align}
    \mathbbm{E}_r[ \braket{Z}_{m_r}^C \braket{Z}_{m_r}^C] = \mathbbm{E}_m[  \braket{Z}_{m}^C \braket{Z}^C_m],
\end{align}
which we refer to as the `classical-classical' object. Agreement between quantum-classical and classical-classical quantities allows one to verify that the quantum simulation behaves, at least at an averaged level, in the same way as the classical simulation. Unfortunately, such a comparison does not yet provide us with information about intrinsic properties of the quantum system, such as the `quantum-quantum' object $\mathbbm{E}_m[  \braket{Z}_m  \braket{Z}_m]$.

However, we can take a step beyond Refs.~\cite{garratt2022measurements,hoke2023quantum,li2021robust} using the trivial inequality $\mathbbm{E}_r[ ( \braket{Z}_{m_r} - \braket{Z}_{m_r}^C)^2] \geq 0$, which can be rewritten using Eq.~\eqref{eq:rmaverage} as
\begin{align}
	\mathbbm{E}_m[\braket{Z}_{m}^2] \geq \mathbbm{E}_r\Big[ 2z_r \braket{Z}^C_{m_r} - \braket{Z}_{m_r}^C  \braket{Z}_{m_r}^C \Big]. \label{eq:trivialbound}
\end{align}
Equation~\eqref{eq:trivialbound} shows us that, although we cannot determine $\mathbbm{E}_m[\braket{Z}_{m}^2]$ directly, we can bound it using cross-correlations between classical and quantum simulations. Before deriving bounds on the measurement-averaged entanglement entropy, we generalize the above scheme using shadow tomography. 

\section{Incorporating shadows}\label{sec:shadows}

It will be useful to first outline shadow tomography \cite{elben2023randomized} in its simplest incarnation for a single qubit. If one can prepare a density matrix $\rho$ multiple times, in each run $r$ of the experiment one applies a random unitary $U_r$ and then measures $Z$, finding result $z_r$. The shadow $\rho^S_r = \rho^S(z_r,U_r)$ is defined as the matrix satisfying
\begin{align}
	\mathbbm{E}_r[\rho^S_r] = \sum_{U,z} P_U P_{z|U} \rho^S(z,U) = \rho,
\label{eq:shadowdef}
\end{align}
where $P_U$ is the probability that we chose the unitary $U$ and $P_{z|U} = \braket{z|U\rho U^{-1}|z}$ is the Born probability for finding result $z$ given that we acted with $U$. If the set of $U$ forms a two-design for the Haar ensemble of single-qubit unitary operators, we should choose
\begin{align}
	\rho^S_r =3 U_r^{-1}\ket{z_r}\bra{z_r}U_r -\mathbbm{1},\label{eq:shadow}
\end{align}
where $Z\ket{z_r}=z_r\ket{z_r}$. Note that $\rho^S_r$ is not a valid density matrix since its eigenvalues are not all positive. It is nevertheless the case that its average over $z$ and $U$ is the density matrix $\rho$. This behavior can be verified using the first two moments of the Haar distribution over the unitary group. Generalizing Eq.~\eqref{eq:shadow} to multiple qubits is straightforward \cite{huang2020predicting}: one possibility is to construct $N$-qubit shadows as the tensor products of $N$ objects with the structure $3 U^{-1}\ket{z}\bra{z}U -\mathbbm{1}$.

Shadow tomography requires that $\rho$ can be prepared multiple times. In our case we cannot efficiently prepare the post-measurement density matrices $\rho_m$ but, as we now show, we can nevertheless apply some of the ideas above. Suppose that in each run $r$ we prepare a single-qubit density matrix $\rho_{m_r}$, choose a random unitary $U_r$, and measure $Z$, finding a result $z_r$. The shadow in this run is given by Eq.~\eqref{eq:shadow} as usual, or as a tensor product of these objects if $\rho_{m_r}$ is a density matrix for multiple qubits. Now we write the matrix-analog of Eq.~\eqref{eq:zdecompose} as
\begin{align}
    \rho^S_r = \rho_{m_r} + [\rho^S_r-\rho_{m_r}]
\end{align}
where $[\rho^S_r - \rho_{m_r}]$ is a mean zero fluctuation. Given an $m$-dependent set of matrices $W_m$, as in Eq.~\eqref{eq:wm} our average over runs can wash out these fluctuations while preserving information on the ensemble of $\rho_m$,
\begin{align}
    \mathbbm{E}_r[W_{m_r} \rho^S_r] = \sum_m P_m W_m \sum_{U,z} P_U P_{z|U,m} \rho^S(z,U),
\end{align}
where $P_{z|U,m}=\braket{z|U \rho_m U^{-1}|z}$. Provided the unitary operations used to construct the shadows are sampled independently of $m$, we can perform the averages over $U$ and $z$ in this expression to arrive at
\begin{align}
	\mathbbm{E}_r[W_{m_r} \rho^S_r] = \mathbbm{E}_m[W_m \rho_m]. \label{eq:averageovernoise}
\end{align}
This relation will allow us to probe quantum-information theoretic quantities without post-selection. 

If we choose $W_m$ based on a classical simulation, for example setting it equal to the classical estimate for the post-measurement density matrix $W_m = \rho^C_m$, we arrive at a matrix-valued cross-correlation $\mathbbm{E}_r[\rho^C_{m_r} \rho^S_r] = \mathbbm{E}_m[\rho^C_m \rho_m]$. Taking the trace of this object would give us a `quantum-classical' purity $\mathbbm{E}_m[\text{Tr}(\rho^C_m \rho_m)]$. In analogy with Eq.~\eqref{eq:trivialbound}, we can use the inequality $\text{Tr}[(\rho_m - \rho^C_m)^2]\geq 0$ to lower bound the measurement-averaged purity
\begin{align}
	\mathbbm{E}_m[\text{Tr} \rho_m^2] \geq \mathbbm{E}_r\Big[ 2\text{Tr}(\rho^C_{m_r} \rho^S_r) - \text{Tr}(\rho^C_{m_r}\rho^C_{m_r} )\Big]. \label{eq:puritybound}
\end{align}
Equation~\eqref{eq:puritybound} is our first bound on an entanglement measure. Note that, if $\rho_m$ is the density matrix of a finite number of qubits, the variance of the right-hand side of this expression is finite.

\section{Entanglement entropy}\label{sec:entanglement}

Here we construct resource-efficient upper [Eq.~\eqref{eq:boundentanglement}] and lower [Eq.~\eqref{eq:doublebound}] bounds on the measurement-averaged von Neumann entanglement entropy. First it is useful to collect some definitions. The entanglement entropy of $\rho_m$ is defined as
\begin{align}
	S_m = -\text{Tr}[\rho_m \log\rho_m], \label{eq:QQ}
\end{align}
and our quantum-classical cross-correlation will come from choosing matrix-valued weights
\begin{align}
	W_m = -\log \rho^C_m,
\end{align}
where $\rho^C_m$ is a classical estimate for the post-measurement density matrix, i.e. $W_m$ is the modular Hamiltonian corresponding to $\rho^C_m$. The estimate $W_m$ could be produced by an approximate classical simulation, or another classical model that takes the measurement outcomes as input (e.g. a machine learning model). In each run $r$ of the experiment we therefore have a single shadow $\rho^S_r$ as well as the calculated quantity $-\log \rho^C_{m_r}$. From these we construct
\begin{align}
	S^{S C}_r = -\text{Tr}[\rho^S_r \log \rho^C_{m_r}], \label{eq:sigmaC}
\end{align}
which, by Eq.~\eqref{eq:averageovernoise}, converges to the quantum-classical entanglement entropy upon averaging over experimental runs
\begin{align}
    \mathbbm{E}_r[S^{S C}_r] = \mathbbm{E}_m[S^{Q C}_m]. \label{eq:SCSQ}
\end{align}
Here we have defined the quantum-classical entanglement entropy
\begin{align}
	S^{Q C}_m = -\text{Tr}[\rho_m \log \rho^C_m]. \label{eq:QC}
\end{align}
In Sec.~\ref{sec:variance} we discuss the number of experimental runs required to observe this convergence. It will often be useful to compare $\mathbbm{E}_m[S^{Q C}_m]$ with the average of the `classical-classical' entanglement entropy
\begin{align}
	S^{CC}_m = -\text{Tr}[\rho^C_m \log \rho^C_m]. \label{eq:CC}
\end{align}
Throughout this work we reserve the name `entanglement entropy' for the object in Eq.~\eqref{eq:QQ}, rather than the proxies in Eqs.~\eqref{eq:sigmaC}-\eqref{eq:CC}. It is important to recognize that, without post-selection, only $S^{S C}_r$ and $S^{CC}_m$ can be determined for individual outcomes $m$. It is the average over measurement outcomes that gives us access also to $\mathbbm{E}_m[S^{Q C}_m]$.

The bound on $\mathbbm{E}_m[S_m]$ comes from writing the quantum relative entropy \cite{nielsen2000quantum} between $\rho_m$ and $\rho^C_m$ as
\begin{align}
	D_m &\equiv \text{Tr}[\rho_m(\log \rho_m - \log \rho^C_m)] \notag \\
 &=S^{QC}_m-S_m, \label{eq:relative}
\end{align}
From the non-negativity of the quantum relative entropy $D_m \geq 0$ we then have
\begin{align}
	\mathbbm{E}_m[S_m] \leq \mathbbm{E}_m[S^{Q C}_m] = \mathbbm{E}_r[S^{SC}_r], \label{eq:boundentanglement}
\end{align}	
where the right-hand side is experimentally accessible. This result shows that, although we cannot determine $\mathbbm{E}_m[S_m]$ without post-selection, we can construct an upper bound using experimental data and classical simulations. Despite its simplicity, Eq.~\eqref{eq:boundentanglement} is one of our central results: the fact that it is possible to bound simulation-independent properties of the quantum system using quantum-classical cross-correlations opens the door to new kinds of experiments in quantum simulators. Interestingly, Eq.~\eqref{eq:boundentanglement} also reveals that we can use experimental data to optimize our models by minimizing $\mathbbm{E}_r[S^{SC}_r]$.

Note that, in order for Eq.~\eqref{eq:boundentanglement} to hold when some of the eigenvalues of $\rho^C_m$ are zero (for example, when $\rho^C_m$ is pure), it is necessary to define $-\log \rho^C_m$ as having infinite contributions, i.e. to define the negative logarithms of the zero eigenvalues as infinity. Since we have the freedom to choose $\rho^C_m$, this is unlikely to be an issue in practice, and we can simply impose that all eigenvalues of $\rho^C_m$ be larger than a threshold.

To arrive at a lower bound on the measurement-averaged entanglement we consider two subregions $A$ and $B$. The true post-measurement density matrices for $AB$ are denoted by $\rho_{AB,m}$ and our classical estimates by $\rho^C_{AB,m}$. It is helpful to first discuss the conditional entropy
\begin{align}
	C_{B|A,m} = S_{AB,m} - S_{A,m}, \label{eq:conditional}
\end{align} 
where $S_{A,m}$ and $S_{AB,m}$ are post-measurement entanglement entropies for $A$ and for $AB$, respectively. For example $S_{A,m} = -\text{Tr}[\rho_{A,m} \log \rho_{A,m}]$. For classical probability distributions the conditional entropy cannot be negative; if it is negative, this is an indication that $A$ and $B$ are entangled. 

Using the definition Eq.~\eqref{eq:relative} we write $S_{A,m}=S^{Q C}_{A,m}-D_{A,m}$, as well as the analogous expression for $S_{AB,m}$. Inserting these relations into $C_{B|A,m}$, we find
\begin{align}
	C_{B|A,m} = C^{QC}_{B|A,m} - D_{AB,m}+ D_{A,m} \leq C^{QC}_{B|A,m}, \label{eq:beforeboundconditional}
\end{align}
where we have defined $C^{QC}_{B|A,m}=S^{Q C}_{AB,m} - S^{Q C}_{A,m}$. The inequality follows from the monotonicity of the relative entropy ${D_{A,m} \leq D_{AB,m}}$. In terms of experimentally-accessible quantities we then have
\begin{align}
	\mathbbm{E}_m[C_{B|A,m}] \leq \mathbbm{E}_r[C^{SC}_{B|A,r}],\label{eq:boundconditional}
\end{align}
where $C^{SC}_{B|A,r}=S^{S C}_{AB,r} - S^{S C}_{A,r}$. Therefore, if we find $\mathbbm{E}_r[C^{SC}_{B|A,r}] \leq 0$, we can provide evidence that $A$ and $B$ are entangled on average.

This bound on the conditional entropy can be recast into a lower bound on the entanglement entropy, complementing the upper bound in Eq.~\eqref{eq:boundentanglement}. Using $C_{B|A,m}\leq C^{QC}_{B|A,m}$ and $S_{AB,m}\geq 0$ we have
\begin{align}
    S^{QC}_{A,m} - S^{QC}_{AB,m} \leq S_{A,m}.
\end{align}
After averaging this can be written as
\begin{align}
   \mathbbm{E}_r[S^{SC}_{A,r}] - \mathbbm{E}_r[S^{SC}_{AB,r}] \leq \mathbbm{E}_m[S_{A,m}] \leq \mathbbm{E}_r[S^{SC}_{A,r}], \label{eq:doublebound}
\end{align}
where on the right-hand side we have also included the upper bound in Eq.~\eqref{eq:boundentanglement} (now with the subregion $A$ of interest specified). The result in Eq.~\eqref{eq:doublebound} shows that the true measurement-averaged von Neumann entanglement entropy can be both lower and upper bounded without post-selection. The upper bound becomes an equality in the limit of a perfect classical simulation. The lower bound, on the other hand, only becomes restrictive for small $\mathbbm{E}_r[S^{SC}_{AB,r}]$, but this is possible only if $\mathbbm{E}_m[S_{AB,m}]$ is itself small. 

Given that we can efficiently constrain the measurement-averaged entanglement, it is natural to ask whether we can say anything about how it changes under quantum operations. Suppose that after preparing $\rho_m$ (for simplicity we here again omit subregion labels) we want to know how its entanglement entropy is modified by a known $m$-independent completely positive trace-preserving (CPTP) linear map, 
\begin{align}
	\rho_m \to \mathcal{E}(\rho_m) \equiv \rho_{m,\mathcal{E}}.
\end{align}
we will denote the entanglement entropy of $\rho_{m,\mathcal{E}}$ by $S_{m,\mathcal{E}}$. An important property of the quantum relative entropy is that it is monotonically decreasing under CPTP maps \cite{lindblad1975completely}, i.e. if $\rho_m \to \rho_{m,\mathcal{E}}$ and $\rho^C_m \to \rho^C_{m,\mathcal{E}} = \mathcal{E}(\rho^C_m)$ then
\begin{align}
	D_{m,\mathcal{E}} \equiv \text{Tr}[\rho_{m,\mathcal{E}} (\log \rho_{m,\mathcal{E}} - \log \rho^C_{m,\mathcal{E}})]
\end{align}
satisfies $D_{m,\mathcal{E}} \leq D_m$. This property implies that, if the map $\mathcal{E}$ increases the measurement-averaged quantum-classical entanglement entropy, it must also increase the true measurement-averaged entanglement entropy:
\begin{align}
	\mathbbm{E}_r[S^{S C}_{r,\mathcal{E}}] \geq \mathbbm{E}_r[S^{S C}_{r}] \implies \mathbbm{E}_m[S_{m,\mathcal{N}}] \geq \mathbbm{E}_m[S_m], \label{eq:changeS}
\end{align}
where we have defined $S^{S C}_{r,\mathcal{E}} \equiv -\text{Tr}[\mathcal{E}(\rho^S_r) \log \mathcal{E}(\rho^C_{m_r})]$, i.e. the channel is applied to the shadow and also to the classical estimate for the post-measurement density matrix $\rho^C_{m_r}$. The quantity $\mathbbm{E}_r[S^{S C}_{r,\mathcal{E}}]$, like $\mathbbm{E}_r[S^{SC}_r]$, can therefore be determined entirely through classical post-processing of the experimentally-determined shadows $\rho^S_r$. 

\section{Statistical fluctuations}\label{sec:variance}

Above we have shown that measurement-averaged entanglement properties can be bounded using particular quantum-classical cross-correlations. These bounds hold provided the averages over runs of the experiment have converged, i.e. if $\mathbbm{E}_r[S^{SC}_r] = \mathbbm{E}_m[S^{QC}_m]$, but in practice the number of runs $R$ is finite. Consequently, these bounds will only be accurate to within an error that decays as $\sim R^{-1/2}$. 

In this section we discuss the statistical fluctuations of $S^{SC}_r$ over runs, focusing for simplicity on the case of a single qubit. This example will illustrate the role played by small eigenvalues of $\rho^C_{m_r}$ in the convergence of the average of $S^{SC}_r$. Some care is required in directly applying bounds on the variance in Ref.~\cite{huang2020predicting} since these diverge as $\rho^C_m$ approaches a pure state.

The mean-squared fluctuation of $S^{SC}_r$ around $S^{QC}_{m_r}$ for a particular outcome $m=m_r$ is
\begin{align}
    (\sigma^{SC}_m)^2 = \sum_{z,U} P_{z,U|m} \text{Tr}\big[\big(\rho^S(z,U)-\rho_m\big) \log \rho^C_m\big]^2, \label{eq:defsigma}
\end{align}
with $\rho^S(z,U)$ given as usual by Eq.~\eqref{eq:shadow} (see above Eq.~\eqref{eq:shadowdef}), and $P_{z,U|m}$ the probability that we acted with $U$ and observed $z$ given outcomes $m$. To calculate $\sigma^{SC}_m$ we require certain properties of the first three moments of the Haar distribution over the unitary group. Haar invariance implies
\begin{align}
\sum_U p_U \prod_{k=0}^{n-1} U_{i_k 0} U^*_{j_k 0} = C_n\sum_{\pi} \prod_{k=0}^{n-1} \delta_{i_k j_{\pi(k)}}, \label{eq:defCn} 
\end{align}
where $\sum_{\pi}$ denotes a sum over all permutations $\pi$ of $n$ elements. By contracting indices in Eq.~\eqref{eq:defCn} it is straightforward to show that for $n\leq 3$ the coefficients $C_n =[(n+1)!]^{-1}$. This expression holds for Haar-random unitary operations as well as those drawn from the Clifford group because the latter forms a three-design \cite{webb2016clifford,zhu2017multiqubit}.

Using Eq.~\eqref{eq:defCn} in Eq.~\eqref{eq:defsigma} we find that $(\sigma^{SC}_m)^2$ can be expressed, up to prefactors, as the sum of three contributions (i) the variance $v_m$ of $-\log \rho^C_m$ with respect to $\rho_m$, (ii) the variance $v_{\mathbbm{1}}$ of $-\log \rho^C_m$ with respect to the maximally-mixed state $\mathbbm{1}/2$, and (iii) the difference $\Delta_{m\mathbbm{1}}$ between the expectation values of $-\log \rho^C_m$ computed with respect to $\rho_m$ and $\mathbbm{1}/2$. In full,
\begin{align}
	(\sigma^{SC}_m)^2 = \frac{3}{2}(v_m + v_{\mathbbm{1}}) + \frac{1}{2}\Delta_{m\mathbbm{1}}^2 \label{eq:VQC}.
\end{align}
where
\begin{align}
	v_{\mathbbm{1}} &= \text{Tr}[ (1/2)(\log \rho^C_m)^2] - \text{Tr}[ (1/2) \log \rho^C_m]^2.\notag \\
	v_m &= \text{Tr}[ \rho_m (\log \rho^C_m)^2] - \text{Tr}[ \rho_m \log \rho^C_m]^2 \label{eq:vm} \\
	\Delta_{m\mathbbm{1}} &= \text{Tr}[(\rho_m-\mathbbm{1}/2)\log \rho^C_m]. \notag
\end{align}
These results make clear that, no matter how accurate our simulation, as $\rho^C_m$ approaches a pure state and hence has eigenvalues approaching zero, the variance $(\sigma^{SC}_m)^2$ diverges. 

Let us imagine that we have access to a perfect classical simulation, $\rho^C_m = \rho_m$, and that our aim is to observe purification of a qubit to within a threshold $0 \leq \mathbbm{E}_m[S_m] \leq \epsilon \log(1/\epsilon)$, for $\epsilon \ll 1$. We have parameterized the threshold in this way since for $\epsilon \ll 1$ it corresponds to $\rho_m$ having minimum eigenvalues of order $\epsilon$. For the variance we then have $(\sigma^{SC}_m)^2 \sim v_{\mathbbm{1}} \sim [\log(1/\epsilon)]^2$. In order to suppress the error in the mean $\mathbbm{E}_r[S^{S C}_r]$ to well below $\epsilon\log(1/\epsilon)$ we therefore require a number of experimental runs $R$ satisfying $R \gg \epsilon^{-2}$. 

The total variance of $S^{SC}_r$ relevant to experiment is the sum of (i) the average of $(\sigma^{SC}_{m_r})^2$ over observed runs $m_r$ and (ii) the variance of $S^{QC}_{m_r}$ over $m_r$. The latter depends sensitively on the ensemble of post-measurement states $\rho_{m_r}$, but crucially it is bounded in the limit of a perfect simulation $\rho^C_m = \rho_m$. This is because we then have $S^{QC}_m=S_m$, and the entanglement entropy $S_m$ of a finite number of qubits is itself finite. 

This section concludes our general discussion of our bounds on the measurement-averaged entanglement entropy. In summary, the average post-measurement entanglement entropy of a finite number of qubits can be upper bounded by a cross-correlation $\mathbbm{E}_r[S^{S C}_r]$ taking as input (i) experimentally-determined shadows $\rho^S_r$ extracted from a quantum simulation and (ii) classical estimates for post-measurement density matrices $\rho^C_{m_r}$. For a perfect classical simulation this upper bound converges to the true measurement-averaged entanglement entropy $\mathbbm{E}_m[S_m]$, which cannot be observed directly without post-selection. Since the quality of the classical simulation controls the quality of our bound, it will be useful to investigate a measurement-induced collective phenomenon that is associated with the breakdown of classical simulations. 

\section{Entanglement transition}\label{sec:transition}
\begin{figure}
	\includegraphics[width=0.46\textwidth]{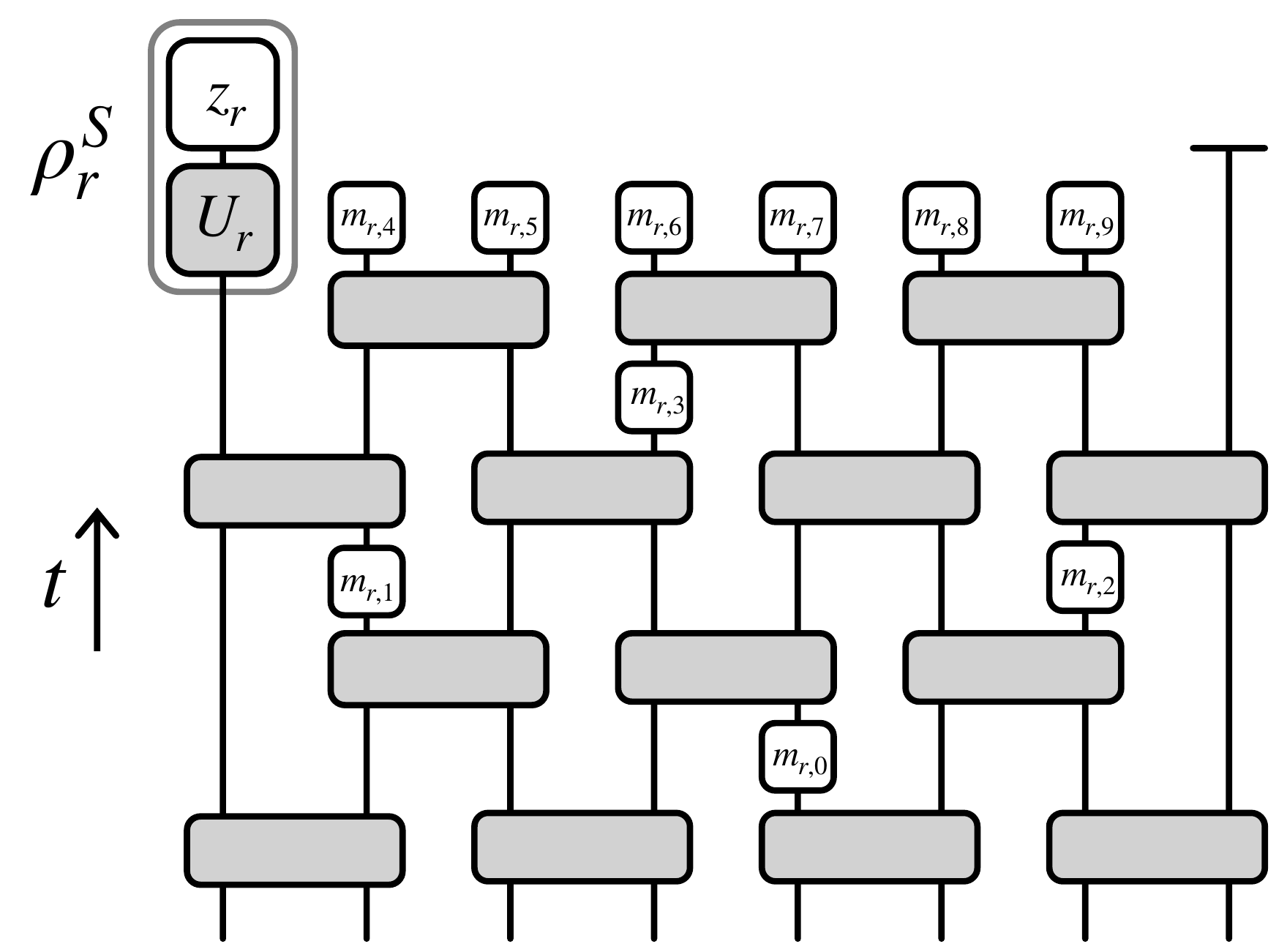}
\caption{Probe of measurement-induced entanglement transition. In each run $r$, $L$ initially unentangled qubits are evolved through a circuit of depth $t$ involving two-qubit Haar-random unitary operations (wide gray boxes) and, with a probability $p$ at each time step, single-qubit measurements of $Z$ (small white boxes, with outcomes $m_{r,0}, \ldots, m_{r,3}$ in the figure). At the final time all bulk qubits are measured (here outcomes $m_{r,4}, \ldots, m_{r,9}$). One then extracts a shadow $\rho^S_r$ from the left boundary qubit by applying a single-qubit random unitary $U_r$ and subsequently measuring $Z$ (with result $z_r$). The outcomes $m_r=(m_{r,0},\ldots,m_{r,9})$ are then used as input to a classical calculation giving an estimate $\rho^C_{m_r}$ for the post-measurement density matrix of the left qubit, and from $\rho^S_r$ and $\rho^C_{m_r}$ we construct $S^{SC}_r$, which is averaged over $r$. For illustration we here show $N=8$ and $t=4$, although for numerical calculations we fix $t=4L$.}
\label{fig:circuit}
\end{figure}

In this section we consider the application of our approach to study the measurement-induced entanglement transition in random quantum circuits \cite{potter2022entanglement,fisher2023random}. Our focus is on one-dimensional chains of qubits evolved under Haar-random two-qubit unitary gates and single-qubit projective measurements \cite{skinner2019measurement}. Unitary gates are arranged in a brickwork pattern and measurements are performed at a fixed rate $p$ per qubit [see Fig.~\ref{fig:circuit} for an illustration]. Previous work on this model has revealed a transition between volume-law entangled states for $p<p_c$ and area-law entangled states for $p>p_c$ \cite{li2018quantum,skinner2019measurement,potter2022entanglement,fisher2023random}, with the critical point estimated as $p_c \simeq 0.16$ \cite{zabalo2020critical}. To clarify notation, $p$ refers to the probability that we choose to measure a qubit, while $P_m$ is the Born probability corresponding to a set of outcomes $m$. 

This entanglement transition is a useful setting to test our idea because it corresponds to the point where classical simulations break down: for $p < p_c$ the classical memory required to describe the state is exponential in the number $L$ of qubits, whereas for $p>p_c$ it is expected that the required memory is only linear in $L$. To emphasize the limitations of classical memory, we will construct cross-correlations between exact numerics (representing the experiment or quantum simulation) and approximate numerics based on MPSs restricted to finite bond dimension $\chi$ (the classical simulation). The classical memory required for the MPS simulation then scales only as $\sim L \chi^2$. We stress that, in this illustrative example, both kinds of simulation are carried out on a classical computer; the phrases `quantum simulation' and `classical simulation' are an allusion to how our idea could be implemented in experiment (see, for example, the related calculation in Ref.~\cite{hoke2023quantum}). 

Care is required when sampling measurement outcomes: while the effects of particular measurement outcomes can be `forced' in a classical simulation (through the application of a desired projection operator), they cannot be forced in experiment. Therefore, here we must sample the outcomes of measurements according to the Born rule in the quantum simulation (i.e. with Born probabilities calculated using exact numerics) and then force these outcomes in the classical simulation.

We choose to work with chains having open boundary conditions rather than periodic since MPS simulations are then much more efficient \cite{schollwock2011density}. Because the measurement-induced entanglement transition appears to have a dynamic critical exponent of unity \cite{skinner2019measurement,li2018quantum} we evolve the system for a time $t=4L$, corresponding to $2L$ two-site unitary operations on each bond. After each such operation we measure the two evolved qubits with a probability $p$, with outcomes sampled as described in the previous paragraph, and with the caveat that during the evolution we choose not to measure the qubits at the left and right ends of the chain. Following the measurement step, we truncate the bond dimension in the MPS simulation to $\chi$. The bond dimension therefore increases from a maximum of $\chi$ to $4\chi$ through the unitary operation, is potentially decreased by measurement, and finally is truncated back to a maximum of $\chi$. At the final time we have an exact representation of the quantum state (from the quantum simulation) and a MPS with bond dimension $\chi$. In the next section we discuss a way to characterize the post-measurement state.

\begin{figure*}
	\includegraphics[width=\textwidth]{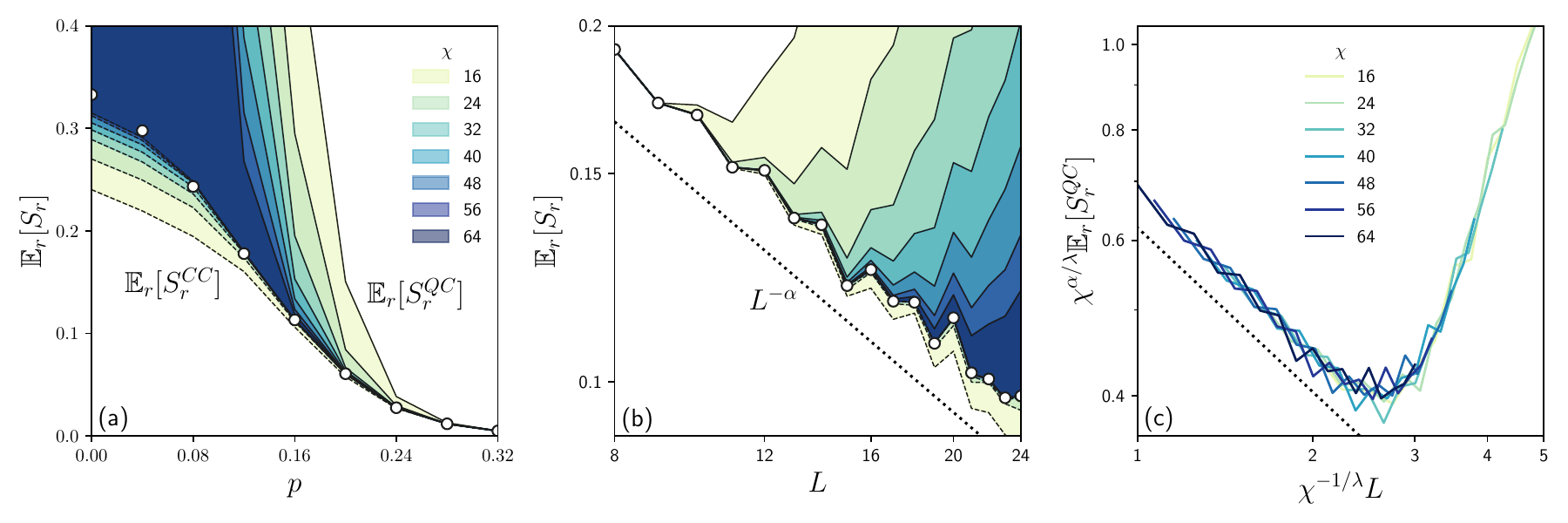}
	\caption{Estimates for average entanglement entropy of an edge site after measuring bulk qubits. (a)  $L=20$ with various $p$ and $\chi$. The different shaded regions, corresponding to different $\chi$, have lower edges $\mathbbm{E}_r[S^{CC}_r]$ (indicated by dashed gray lines) and upper edges $\mathbbm{E}_r[S^{Q C}_m]$ (indicated by solid gray lines). Narrower shaded regions correspond to larger $\chi$ (see legend). The open circles correspond to the exact average entropy $\mathbbm{E}_r[S_r]$. (b) Entanglement in the critical regime: $p=0.16$ and for various system sizes $L$. The dotted black line shows a power-law fit $\sim L^{-\alpha}$ to $\mathbbm{E}_r[S_r]$ (open circles). The exponent $\alpha = 0.62 \pm 0.02$, and the fit is offset from the data for clarity. (c) Scaling collapse of $\mathbbm{E}_r[S^{QC}_r]$, corresponding to upper edges of shaded regions in (b), for various $\chi$. For this collapse we use $\lambda=2$, and we discuss this exponent in detail in Sec.~\ref{sec:memory}. The dotted line shows decay $\sim L^{-\alpha}$ as in (b).}
	\label{fig:bigfig}
\end{figure*}

\subsection{Upper bound}\label{sec:upper}
To probe quantum correlations we must focus on density matrices of small numbers of qubits. This is because the variance of $S^{SC}_r$ grows exponentially with the number of qubits being investigated \cite{huang2020predicting}. The local probe that we choose is as follows: at ${t=4L}$ we measure all but the left- and right-most qubits in the chain, and we then try to estimate the entanglement entropy of the left qubit (this is equal to the entanglement entropy of the right qubit since the overall state is pure); our protocol is represented in Fig.~\ref{fig:circuit}. 

Our numerical (Monte Carlo) average over runs $\mathbbm{E}_r[\cdots]$ will in this section also include an average over unitary gates used to construct the quantum circuit. Since the post-measurement density matrices of the left-qubit are determined not only by outcomes $m_r$ observed in run $r$, but also by the unitary gates in that run, we will here use the more general notation $\rho_r$ rather than $\rho_{m_r}$. It is important to note that $m_r$ represents the list of outcomes recorded during evolution and also at the final time. In this notation, the classical estimate for $\rho_r$ in run $r$ is $\rho^C_r$, and the shadow is $\rho^S_r$, with e.g. $S_r=-\text{Tr}[\rho_r \log \rho_r]$ and $S^{SC}_r=-\text{Tr}[\rho^S_r \log \rho^C_r]$. The bound in Eq.~\eqref{eq:boundentanglement} is then written ${\mathbbm{E}_r[S_r] \leq \mathbbm{E}_r[S^{QC}_r] = \mathbbm{E}_r[S^{SC}_r]}$. Since we are here using exact numerics in place of an actual quantum simulation, for numerical convenience we directly determine $\rho_r$ and then average $S^{Q C}_r$ rather than $S^{S C}_r$. The result is the same for a large number of runs.

A reason for considering the post-measurement entanglement between the boundary qubits in the present setting is that it is straightforward to understand some of the effects of a finite bond dimension. For any $\chi$, Eq.~\eqref{eq:boundentanglement} guarantees that $\mathbbm{E}_r[S^{Q C}_r] \geq \mathbbm{E}_r[S_r]$, with the difference expected to increase with $L$. On the other hand, for finite $\chi$ we can compute $\rho^C_m$ as a product of essentially random finite-dimensional matrices; without fine tuning there is a gap between the two leading Lyapunov exponents characterizing this product. As a consequence, $\mathbbm{E}_r[S^{CC}_r]$ must decay exponentially with $L$ for any finite $\chi$. Since $\mathbbm{E}_r[S_r]$ is $L$-independent for $p<p_c$ and decays as a power of $L$ for $p=p_c$ we see that, at fixed $\chi$ and for sufficiently large $L$, the classical estimate for the measurement-averaged entanglement $\mathbbm{E}_r[S^{CC}_r]$ should lower bound the true entanglement entropy $\mathbbm{E}_r[S_r]$. Having an approximate lower bound of this kind is useful because, by increasing $\chi$, we can expect the chain of inequalities $\mathbbm{E}_r[S^{CC}_r] \lesssim \mathbbm{E}_r[S_r] \leq \mathbbm{E}_r[S^{QC}_r]$ to become more restrictive. We defer a discussion of the rigorous lower bound in Eq.~\eqref{eq:doublebound} to Sec.~\ref{sec:lower}.

First consider sweeping across the transition. In Fig.~\ref{fig:bigfig}(a) we show $\mathbbm{E}_r[S_r]$,  $\mathbbm{E}_r[S^{QC}_r]$ and $\mathbbm{E}_r[S^{CC}_r]$ for $L=20$ and various $\chi$. On increasing $\chi$, the window between $\mathbbm{E}_r[S^{CC}_r]$ and $\mathbbm{E}_r[S^{QC}_r]$ becomes narrower and, as discussed above, $\mathbbm{E}_r[S_r]$ should lie within this window at large $L$. Interestingly, $\mathbbm{E}_r[S^{Q C}_r]$ departs from the true measurement-averaged entanglement entropy at a much larger value of $p$ than $\mathbbm{E}_r[S^{CC}_r]$. The splitting of $\mathbbm{E}_r[S^{Q C}_r]$ and $\mathbbm{E}_r[S^{CC}_r]$ signifies the breakdown of the simulation: the apparent agreement between $\mathbbm{E}_r[S^{CC}_r]$ and $\mathbbm{E}_r[S_r]$ even for small $p$ is an artefact of the ensemble average. This is to say that, although $\mathbbm{E}_r[S^{CC}_r]$ and $\mathbbm{E}_r[S_r]$ appear to agree, we know that $S^{CC}_r$ and $S_r$ disagree for individual runs $r$ because $\mathbbm{E}_r[S^{QC}_r]$ and $\mathbbm{E}_r[S_r]$ disagree. 

Turning now to the critical regime $p=0.16$, in Fig.~\ref{fig:bigfig}(b) we investigate the effect of increasing $L$ for various $\chi$. As expected, the decay of $\mathbbm{E}_r[S_r]$ is consistent with a power-law $\sim L^{-\alpha}$, and we find numerically that $\alpha \approx 0.6$. For each $\chi$ we see that, as $L$ is increased, there is a point beyond which $\mathbbm{E}_r[S^{Q C}_r]$ deviates from this power law decay, and instead begins to increase. 

It is natural to ask how the system size $L$ at which this occurs depends on $\chi$. This dependence will control the classical memory required to observe measurement-induced criticality in experiment. In Fig.~\ref{fig:bigfig}(c) we show that $\mathbbm{E}_r[S^{Q C}_r]$ can be collapsed according to the scaling form
\begin{align}
    \mathbbm{E}_r[S^{Q C}_r] \simeq A L^{-\alpha} f(\chi L^{-\lambda}), \label{eq:collapse}
\end{align}
where $A$ is the prefactor of the $L^{-\alpha}$ decay in $\mathbbm{E}_r[S_r]$, and $f(x)$ is a function satisfying $f(x \to 1)=1$ for large $x$ (and which is large at small $x$). In Fig.~\ref{fig:bigfig}(c) we have set the exponent $\lambda = 2$, but we discuss this further below. This collapse is an experimentally observable signature of measurement-induced criticality, which can be obtained without post-selection through the average $\mathbbm{E}_r[S^{SC}_r]$. We discuss this signature, and the resources required to observe it, in the next section.

\subsection{Resource requirements}\label{sec:memory}

The classical memory required to store a MPS scales as $\sim L \chi^2$. For $p < p_c$ we expect that the bond dimension required to obtain a reasonable approximation to the true quantum state is exponential in $L$, so determining the measurement-averaged entanglement there is presumably a lost cause. At the critical point $p=p_c$ itself, Fig.~\ref{fig:bigfig}(c) suggests that with bond dimension $\chi \gtrsim L^{\lambda}$ we can find a restrictive bound on $\mathbbm{E}_r[S_r]$. This is plausible because, prior to measuring all but the left and right qubits, the half-chain entanglement entropy at $p=p_c$ is proportional to $\log L$ \cite{potter2022entanglement,fisher2023random}. Since the half-chain entanglement is upper bounded by $\log \chi$, one would only need a bond dimension polynomial in $L$ to capture this. 

\begin{figure}
\includegraphics[width=0.47\textwidth]{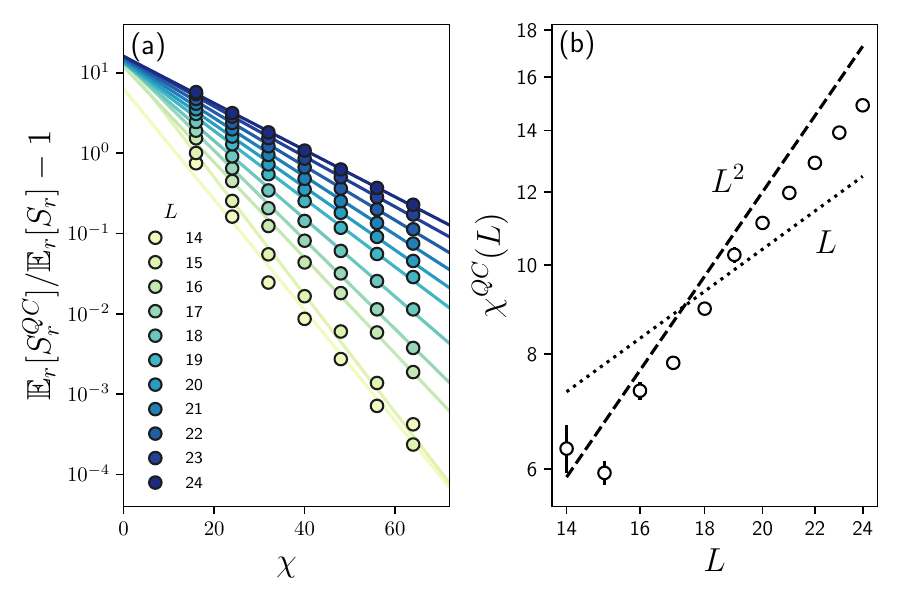}
\caption{Effect of varying bond dimension $\chi$ at $p=0.16$. (a) Decay of $\mathbbm{E}_r[S^{Q C}_r]/\mathbbm{E}_r[S_r]-1$ with $\chi$ for various $L$ (open circles), and fits to exponential decays (solid lines). The decay rates give the parameter $\chi^{QC}(L)$ in Eq.~\eqref{eq:chiscaling}. (b) Growth of $\chi^{QC}(L)$ with $L$ (open circles), with logarithmic scaling of both axes, compared with power laws $L$ (dotted) and $L^2$ (dashed).}
\label{fig:chi}
\end{figure}

To explore the $\chi$ dependence of $\mathbbm{E}_r[S^{Q C}_r]$ in detail, in Fig.~\ref{fig:chi}(a) we calculate $\mathbbm{E}_r[S^{Q C}_r]/\mathbbm{E}_r[S_r]$. There we find that the convergence of $\mathbbm{E}_r[S^{Q C}_r]$ is exponential
\begin{align}
	\mathbbm{E}_r[S^{Q C}_r]/\mathbbm{E}_r[S_m]-1 \propto e^{-\chi/\chi^{QC}(L)},\label{eq:chiscaling}
\end{align}
with a constant of proportionality that is of order unity and approximately $L$-independent. It can be verified that this is consistent with Fig.~\ref{fig:chi}(b) provided $\chi$ is not too much smaller than $\chi^{QC}(L)$. The behavior in Fig.~\ref{fig:chi}(a) provides further evidence that there is a characteristic bond dimension $\chi^{QC}(L)$ beyond which $S^{QC}_r$ and $S_r$ are in good agreement, and that this bond dimension increases with $L$. Following Eq.~\eqref{eq:collapse} we expect $\chi^{QC}(L) \sim L^{\lambda}$ and in Fig.~\ref{fig:chi}(b) we show $\chi^{QC}(L)$ as a function of $L$. Our results are consistent with a power-law growth having $\lambda \lesssim 2$; a more accurate determination of $\lambda$ will require access to larger system sizes. Combining Eqs.~\eqref{eq:collapse} and \eqref{eq:chiscaling} we see that the function $f(x)$ takes the form
\begin{align}
    f(x) = 1 + B e^{-x/x_0}, \label{eq:fx}
\end{align}
where $B$ and $x_0$ are constants. For the particular model that we consider, we find from Fig.~\ref{fig:chi} that $B \approx 15$ while $x_0 \approx 0.04$. The observation that $\lambda \lesssim 2$ implies that a good MPS representation of the quantum state need only occupy polynomial classical memory ${L \chi^2 \sim L^{1+2 \lambda} \lesssim L^5}$. 

The power-law scaling of the bond-dimension with a length scale echoes the theory of finite entanglement scaling at conventional quantum critical points \cite{tagliacozzo2008scaling,pollmann2009theory,pirvu2012matrix}. This theory predicts that a finite bond dimension $\chi$ introduces a finite correlation length $\xi \sim \chi^{\kappa}$ with an exponent $\kappa$ that is a function of the central charge of the conformal field theory describing the critical point. That theory addresses the question of how much classical memory is required to study quantum criticality on a classical computer, and it would be interesting to develop an analogous theory for measurement-induced criticality. In our notation, this would correspond to a theory for the convergence of $\mathbbm{E}_r[S^{CC}_r]$ to $\mathbbm{E}_r[S_r]$.

Our results above, on the convergence of $\mathbbm{E}_r[S^{QC}_r]$ to $\mathbbm{E}_r[S_r]$, address the quite different question of how much classical memory is required to observe measurement-induced criticality in a quantum simulator. The interpretation of the measurement-induced entanglement transition as a kind of replica-symmetry-breaking transition \cite{bao2020theory,jian2020measurement} suggests that the classical memory requirements are different in the two cases: writing $S^{QC}_r$ as the $n \to 1$ limit of the object $[1-n]^{-1} \log \text{Tr}[ \rho_r (\rho^C_r)^{n-1}]$, it is clear that since $\rho_r \neq \rho^C_r$ the symmetry under exchange of replicas is reduced relative to $[1-n]^{-1} \log \text{Tr}[(\rho^C_r)^n]$, which reproduces $S^{CC}_r$ as $n \to 1$. In others words, the constraint of finite $\chi$ is a replica-symmetric perturbation in $S^{CC}_r$ but not in $S^{QC}_r$. 

Having discussed the classical memory requirements, we now turn to the time requirements. This is a question of how many experimental runs are required in order to wash out statistical fluctuations in $S^{S C}_r$. The variance of $S^{SC}_r$ over runs is the sum of two contributions: the average of the variance over shadows $\rho^S_r$ for each $\rho_r$ and $\rho^C_r$, which is represented in Eq.~\eqref{eq:VQC}, and the variance of $S^{QC}_r$ over runs. Numerically, we find that the first of these two contributions dominates, although here we simply calculate the overall variance.

\begin{figure}
	\includegraphics[width=0.47\textwidth]{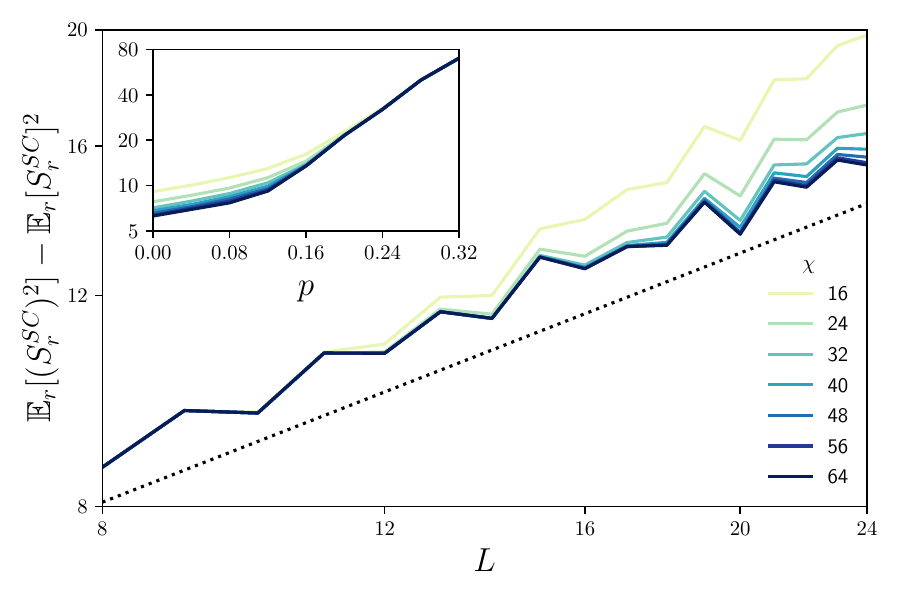}
	\caption{Variance of $S^{SC}_r$ over runs. The main panel shows (with log-log scale) the variance as a function of $L$ for $p=0.16$ and various $\chi$ (see legend). The dotted line shows a fit $\sim L^{\beta}$ to data for $\chi=64$, with $\beta = 0.52 \pm 0.02$, and this is offset from the data for clarity. The inset shows the variance as a function of $p$ for $L=20$ and for the same values of $\chi$ as in the main panel (here with log-linear scale).} 
	\label{fig:var}
\end{figure}

In the main panel of Fig.~\ref{fig:var} we study the growth of the variance with $L$ at $p=0.16$ (in the critical regime). The data suggests that the growth is no faster than ${\sim L^{\beta}}$ for an exponent $\beta > 0$, and a fit to the $\chi=64$ data indicates that $\beta \approx 0.5$ (see Fig.~\ref{fig:var} caption). As in Fig.~\ref{fig:bigfig}(b), as $L$ is increased at fixed $\chi$ there is a point at which the simulation appears to break down. In Fig.~\ref{fig:var} this breakdown manifests as an increase in the variance with decreasing $\chi$. Before this point, where $\chi \gg \chi^{QC}(L)$, our results in Figs.~\ref{fig:bigfig}(b) and \ref{fig:var} suggest that the number of experimental runs required to suppress the error to below the scale of the mean $\mathbbm{E}_r[S_r] \sim L^{-\alpha}$ grows only as $L^{\beta + 2\alpha} < L^2$. 

This result shows that the transition in the behavior of the measurement-averaged von Neumann entanglement entropy can be observed using a number of runs that is polynomial in the system size. However, it is important to note that if one instead aims to bound the measurement-averaged purity, the statistical fluctuations are substantially smaller. This is because, as discussed below Eq.~\eqref{eq:puritybound}, the variance of the object which lower bounds $\mathbbm{E}_r[\rho_r^2]$ is finite. 

\subsection{Lower bound}\label{sec:lower}

For simplicity our focus has been on the upper bound on $\mathbbm{E}_r[S_r]$ in Eq.~\eqref{eq:boundentanglement}, and we have not studied the lower bound in Eq.~\eqref{eq:doublebound} numerically. In order to use Eq.~\eqref{eq:doublebound} it would be necessary to consider a situation where the overall state of the left and right boundary qubits is not quite pure; if it is pure, then for an imperfect classical simulation $\mathbbm{E}_r[S^{QC}_r]$ diverges. 

A simple way to avoid this issue is to apply weak depolarizing channels to $\rho^S_r$ and $\rho^C_r$ in classical post-processing, as described in connection with Eq.~\eqref{eq:changeS}. It will be convenient to use the notation of Eq.~\eqref{eq:doublebound}, and so we denote by $A$ and $B$ the left and right boundary qubits, respectively. If we apply these `artificial' depolarizing channels to $\rho^S_r=\rho^S_{AB,r}$ and $\rho^C_r=\rho^C_{AB,r}$ then
\begin{align}
    \rho^S_{AB,r} &\to \mathcal{E}(\rho^S_{AB,r}) = (1-\epsilon)\rho^S_{AB,r}+(\epsilon/4) \mathbbm{1}, \\
    \rho^C_{AB,r} &\to \mathcal{E}(\rho^C_{AB,r}) = (1-\epsilon)\rho^C_{AB,r}+(\epsilon/4) \mathbbm{1}.\notag
\end{align}
For any finite $\epsilon > 0$, all of the eigenvalues of $\mathcal{E}(\rho^C_{AB,r})$ are nonzero, and hence $\mathbbm{E}_r[S^{SC}_{r,AB,\mathcal{E}}]$ must be finite. Here the notation $S^{SC}_{r,AB,\mathcal{E}}$ indicates that we are considering properties of the density matrices for $AB$ in run $r$ after applying the channel $\mathcal{E}$ in post-processing, while $\mathbbm{E}_r[S_{r,AB,\mathcal{E}}]$ corresponds to the entanglement of the true density matrices had we applied $\mathcal{E}$ in experiment.

Equation~\eqref{eq:doublebound} tells us $\mathbbm{E}_r[S_{r,A,\mathcal{E}}]$ can be determined (rather than just upper bounded) to within a threshold $\mathbbm{E}_r[S^{SC}_{r,AB,\mathcal{E}}]$, so it is pertinent to ask how restrictive we can make this condition. Moreover, as $\epsilon \to 0$, $\mathbbm{E}_r[S_{r,A,\mathcal{E}}] \to \mathbbm{E}_r[S_{r,A}]$. Since the smallest eigenvalues of $\mathcal{E}[\rho_{r,AB}]$ are of order $\epsilon$, we anticipate $\mathbbm{E}_r[S_{r,AB,\mathcal{E}}] \sim \epsilon \log(1/\epsilon)$, and Eq.~\eqref{eq:boundentanglement} implies that this is a lower bound on $\mathbbm{E}_r[S^{SC}_{r,AB,\mathcal{E}}]$. Therefore, in the idealized case of a perfect classical simulation, we would be able to determine $\mathbbm{E}_r[S_{r,A,\mathcal{E}}]$ to within $\sim \epsilon \log(1/\epsilon)$. It would appear at first that we can simply decrease $\epsilon$, but even for a perfect classical simulation the variance of $S^{SC}_{r,AB,\mathcal{E}}$ increases as $\sim [\log(1/\epsilon)]^2$ for small $\epsilon$ (this follows from a generalization of the results in Sec.~\ref{sec:variance} to two-qubit density matrices), limiting the precision with which we can determine $\mathbbm{E}_r[S_{r,A,\mathcal{E}}]$. 

In reality our classical simulations are approximations to the true quantum system. There is then a second contribution to $\mathbbm{E}_r[S^{SC}_{r,AB,\mathcal{E}}]$ scaling as $\sim g(\chi)\log(1/\epsilon)$, where $g(\chi)$ quantifies the mismatch between the eigenvectors of $\rho_{AB,r}$ and $\rho^C_{AB,r}$. If at criticality $g(\chi)$ behaves similarly to $f(\chi L^{-\lambda})$ [Eq.~\eqref{eq:fx}] we should anticipate that this second contribution is subleading beyond $\chi \gg \chi^{QC}(L)\log(1/\epsilon)$. This argument indicates that, for $p = p_c$, the minimum $\epsilon$ we can achieve will ultimately be constrained by classical memory. Investigating the lower bound on the measurement-averaged entanglement numerically is an interesting problem for the future.

Through this section we have studied the quantum-classical entanglement entropy in the context of measurement-induced criticality. In Fig.~\ref{fig:bigfig}(c) we have identified an observable signature of generic measurement-induced criticality that does not require post-selection, and which describes how classical simulations break down; the time and memory required to observe this signature appear to be polynomial in $L$. 

\section{Decoding protocol}\label{sec:decoding}
It is important to understand the relation between $\mathbbm{E}_r[S^{S C}_r]$ and the probe of Ref.~\cite{gullans2020scalable}. For the sake of simplicity, in this section and the next we can consider a protocol in which the only randomness is in the measurement outcomes, and so we return to the notational convention preceding Sec.~\ref{sec:transition}. In Ref.~\cite{gullans2020scalable} the authors consider a scenario where the effects of many measurements, having outcomes $m$, are encoded in the state $\rho_{m}$ of a reference qubit. A generalization of their decoding scheme to the situation where one has an approximate classical simulation of the model is as follows: given an observed set of measurement outcomes $m_r$ in run $r$ one (i) estimates the post-measurement density matrix $\rho^C_{m_r}$, (ii) constructs a unitary decoder $V^C_{m_r}$ such that the Bloch vector of $V^C_{m_r} \rho_{m_r}^C [V^C_{m_r}]^{-1}$ is parallel to the $+Z$ axis, (iii) applies the decoder to the true post-measurement density matrix $\rho_{m_r}$, (iv) measures $Z$ and, (v) averages the results $z_r$ over runs of the experiment. If $\rho_{m_r}$ is pure and the calculation on a classical computer yielding $\rho^C_m$ is perfect, the result of the measurement in step (iv) is $z_r=1$ with certainty. If $\rho_m$ is maximally mixed, the result is $z_r=\pm 1$ with equal probability. 

This approach naturally gives a bound on a measurement-averaged property of the density matrix, as we now discuss, although this bound is distinct from the one in our Eq.~\eqref{eq:boundentanglement}. It will be convenient to define $\gamma_m$ and $\gamma^{CC}_m$ by $\text{Tr}[\rho_m^2] = \frac{1}{2}[1+(\gamma_m)^2]$ and $\text{Tr}[(\rho^C_m)^2] = \frac{1}{2}[1+(\gamma^{CC}_m)^2]$, respectively, where $0 \leq \gamma_m, \gamma^{CC}_m \leq 1$. Note that we then have $V^C_m \rho_m^C [V^C_m]^{-1} = \frac{1}{2}(1+\gamma^{CC}_m Z)$, with $\frac{1}{2}(1+\gamma^{CC}_m)$ the classical estimate for the probability that the $Z$ measurement return $z_r=1$. If we apply the decoder $V^C_m$ to $\rho_m$ and then measure $Z$, the probability to find $z_r=1$ is instead
\begin{align}
	\frac{1}{2}(1+\gamma_m^{QC}) \equiv \braket{1|V^C_m \rho_m [V^C_m]^{-1}|1}, \label{eq:gammaQC}
\end{align}
which defines the quantum-classical object $\gamma^{QC}_m$ satisfying $\gamma_m^{QC} \leq \gamma_m$. Here equality is achieved when the Bloch vectors of $\rho^C_m$ and $\rho_m$ are parallel (but not necessarily equal). This is because an imperfect decoder will generate a density matrix $V^C_m \rho_m^C [V^C_m]^{-1}$ whose Bloch vector is canted relative to the $+ Z$ axis, and hence a measurement of $Z$ has a lower probability to give result $z_r=1$. If we repeat the experiment many times, averaging $z_r$ over runs, the result is
\begin{align} 
	\mathbbm{E}_r[z_r] = \mathbbm{E}_m[\gamma_m^{QC}] \leq \mathbbm{E}_m[\gamma_m]. \label{eq:gammabound}
\end{align} 
Although Eq.~\eqref{eq:boundentanglement} does bound a measurement-averaged property of the ensemble of $\rho_m$, it does not provide us with information on a standard entanglement measure.

On a practical level, the scheme that we presented earlier in this work is based on classical post-processing rather than the application of a conditional decoding unitary $V^{C}_m$. The former approach is simpler in experimental settings, where decoherence timescales may be too short for one to determine and then apply a measurement-conditioned unitary. That said, as we now show, the quantity $\mathbbm{E}_m[\gamma^{QC}_m]$ that is determined via the scheme in Ref.~\cite{gullans2020scalable} can also be determined simply by classical post-processing, and so we expect that if either one is experimentally viable then both should be. 

To see how to determine $\mathbbm{E}_m[\gamma^{QC}_m]$ through classical post-processing we need only consider the average of Eq.~\eqref{eq:gammaQC} over outcomes. In run $r$, corresponding to outcome $m_r$, we extract a shadow $\rho^S_r$ and calculate a matrix-valued weight
\begin{align}
    W_{m_r} = 2[V^C_{m_r}]^{-1} \ket{1}\bra{1} V^C_{m_r}-\mathbbm{1}.
\end{align}
For this $W_{m_r}$ we have 
\begin{align}
	\mathbbm{E}_r[\rho^S_r W_{m_r}] = \mathbbm{E}_m[\gamma_m^{QC}]. \label{eq:gullansshadows}
\end{align}
A different quantum-classical cross-correlation, sharing some features with Eq.~\eqref{eq:gullansshadows} but based on scalar- rather than matrix-valued cross-correlations, was recently implemented in a quantum simulator \cite{hoke2023quantum}. In the notation above, that approach corresponds to replacing $W_{m_r}$ with $\text{sign}(\text{Tr}[\rho^C_{m_r} Z]) Z$.

Inverting this discussion, we can also ask what would be the analog of our scheme for determining $\mathbbm{E}_r[S^{QC}_r]$ if we were to use measurement-conditioned unitary decoders rather than classical post-processing. A hint comes from the fact that we can interpret $S^{QC}_m$ as the expectation value of the $m$-dependent observable $-\log \rho^C_m$ with respect to the true density matrix $\rho_m$. In fact, the protocol would be almost identical to that in Ref.~\cite{gullans2020scalable}, with only a minor difference in the step (v) indicated at the start of this section. That is, having extracted the measurement outcomes $m_r$ in each run $r$, used them as inputs to calculations giving $\rho^C_{m_r}$ and $V^C_{m_r}$, applied $V^C_{m_r}$ to $\rho_{m_r}$, and measured $Z$ (all before the qubits decohere), the only difference is in the number that we should average. If instead of averaging eigenvalues $z_r = \pm 1$ of $Z$, which gives $\mathbbm{E}_r[z_r] = \mathbbm{E}_m[\gamma^{QC}_m]$ in the limit of a large number of runs, one instead averages the corresponding eigenvalues $-\log[(1 + z_r\gamma^{CC}_{m_r})/2]$ of the negative logarithm of the density matrix $\rho_{m_r}^C$, the result converges to $\mathbbm{E}_r[S^{QC}_r]$. 

While the measurement-averaged results from a decoding scheme giving $\mathbbm{E}_m[S^{QC}_m]$ are the same as from classical post-processing using shadows, the statistical fluctuations are likely to be much smaller. This is because, if $\rho^C_m$ is a good estimate for $\rho_m$, in the decoding scheme one is less likely to be affected by small eigenvalues of $\rho^C_m$. This decrease in the scale of statistical fluctuations comes at the cost of having to apply a measurement-conditioned unitary operation. A further issue with a decoding scheme is that it is unclear how to generalize it to study post-measurement density matrices of multiple qubits without introducing multi-qubit unitary operations. Within a shadows scheme involving only classical post-processing the extension is immediate: one can construct multi-qubit shadows simply as tensor products of single-qubit shadows.

\section{Entanglement negativity}\label{sec:negativity}
Our focus in this work has been on bounding the measurement-averaged von Neumann entanglement entropy. While the bounds in Eq.~\eqref{eq:doublebound} hold also in the presence of noise, i.e. when we have access only to a mixed state, the von Neumann entanglement is then no longer a sensitive probe of quantum correlations. In particular, it can be nonzero even when the overall density matrix is diagonal. 

This concern motivates the question of whether it is possible to bound measures of mixed state entanglement. A prominent probe is the entanglement negativity \cite{vidal2002computable}, which can be defined for the density matrix $\rho_{AB}$ of a bipartite system $AB$ as
\begin{align}
	\mathcal{N}(\rho_{AB}) = -\text{Tr}[ P^{(-)}(\rho^{T_A}_{AB}) \rho^{T_A}_{AB}].
\end{align}
Here $\rho^{T_A}_{AB}$ denotes the partial transpose of the density matrix for subsystem $A$, and $P^{(-)}(\rho^{T_A}_{AB})$ is the projector onto the space spanned by the eigenvectors of $\rho^{T_A}_{AB}$ with negative eigenvalues. The negativity has the appealing properties of vanishing for unentangled states and not increasing under local operations and classical communication. Observe now that we can bound the average of the post-measurement negativity $\mathcal{N}(\rho_{m,AB}) = -\text{Tr}[ P^{(-)}(\rho^{T_A}_{m,AB}) \rho^{T_A}_{m,AB}]$ using
\begin{align}
	\mathbbm{E}_m [\mathcal{N}(\rho_{m,AB})] \geq -\mathbbm{E}_r [ P^{(-)}(\rho^{C,T_A}_{m_r,AB}) \rho^{S,T_A}_{r,AB}]. \label{eq:negativitybound}
\end{align}
Here $\rho^{C,T_A}_{m_r,AB}$ is the partial transpose of the classical estimate for the density matrix of $AB$ given outcomes $m_r$, which is used to construct the projector $P^{(-)}(\rho^{C,T_A}_{m_r,AB})$, and $\rho^{S,T_A}_{r,AB}$ is the partial transpose of the shadow in run $r$. Equation~\eqref{eq:negativitybound} shows that we can lower bound the measurement-averaged entanglement negativity by maximizing the right-hand side over projection operators. 

\section{Discussion and outlook}\label{sec:discussion}

The randomness of the quantum measurement process prevents us from directly observing the effects of many measurements in experiment without an exponential post-selection overhead. Cross-correlations (and cross entropies) between quantum experiments and classical simulations \cite{boixo2018characterizing,li2021robust,li2022cross,garratt2022measurements,lee2022decoding,hoke2023quantum} provide a way to avoid this overhead, but in general do not provide information on intrinsic properties of post-measurement quantum states. Moreover, there appeared to be a degree of arbitrariness in exactly how to use information from classical simulations. By showing that certain cross-correlations bound measurement-averaged entanglement measures, the present work has resolved these problems. 

In order to arrive at these bounds we leveraged ideas from shadow tomography \cite{elben2019statistical} to generalize the scalar cross-correlations of Refs.~\cite{li2021robust,garratt2022measurements,lee2022decoding,hoke2023quantum} to matrix-valued quantities. These gave us access to quantum information-theoretic objects, and allowed us to construct a contribution to the quantum relative entropy that can be obtained without post-selection. From properties of the quantum relative entropy we then arrived at both upper and lower bounds on post-measurement entanglement; related cross-correlations were shown to lower bound the purity and negativity. At a minimum, our randomized measurement protocol requires only that individual qubits can be measured in any of the Pauli $X$, $Y$ or $Z$ bases. From each run $r$ of the experiment one extracts a set of measurement outcomes $m_r$ and a shadow $\rho^S_r$ of a finite number of qubits. Then, in post-processing, these shadows are cross-correlated with the results of simulations on classical computers. 

The upper bound in Eq.~\eqref{eq:boundentanglement} can be applied quite generally, and the only restriction is that the subregion of interest should consist of a small finite number of qubits. This is because statistical fluctuations of $S^{SC}_{A,r}$ grow rapidly with the support of $A$ (as in more conventional shadow-tomographic probes). The lower bound in Eq.~\eqref{eq:doublebound}, on the other hand, is only restrictive when we can identify another subregion $B$ such that $AB$ is almost pure. The quantum-classical object $\mathbbm{E}_r[S^{SC}_{AB,r}] \geq \mathbbm{E}_m[S_{AB,m}]$ can then be made small, and hence the bounds in Eq.~\eqref{eq:doublebound} can be made restrictive. This constraint provides strong motivation for experimental protocols of the kind discussed in e.g. Refs.~\cite{bao2022finite,lin2022probing} and our Sec.~\ref{sec:transition}, where the effects of a large number of measurements are ultimately encoded in the entanglement between just two qubits.

Interestingly, our bounds suggest that we can view the post-selection problem as a variational problem, with $\mathbbm{E}_r[S^{SC}_r]$ an objective function taking experimental data as input. For example, if the classical simulation has a parameter (or parameters) $\mu$, then by minimizing $\mathbbm{E}_r[S^{S C}_r]$ one can optimize the bound on $\mathbbm{E}_m[S_m]$, thereby improving the model from experimental data. There is an important pitfall in constructing such a variational scheme, in that $\mu$ must be chosen independently of the random unitary operations $U_r$ and random measurement results $z_r$ used to construct the shadows. Otherwise, the average of $S^{S C}_r$ over runs is not guaranteed to converge to the average of $S^{Q C}_m$ over outcomes. A straightforward way to implement a valid variational scheme is to choose $\mu$ based on one set of experimental runs and then to compute the average of $S^{S C}_r$ using this $\mu$ and a distinct set of runs. In the context of machine learning (see  Ref.~\cite{dehghani2022neuralnetwork} for a different application of machine-learning techniques in the present context) this corresponds to a division into training and test data. 

When applying these ideas to data obtained in current quantum simulators, it is important to account for the presence of noise. 
Significant statistical errors arise when $\rho^C_m$ has small eigenvalues [see for example the behavior at large $p$ in the inset of Fig.~\ref{fig:var}]. Very recently, Ref.~\cite{vermersch2023enhanced} has shown that the idea of common random numbers can be generalized to shadow estimation; this procedure reduces statistical errors provided one has access to a classical estimate for the shadow, which can itself be estimated from a classical estimate for the density matrix. In our case, the scheme corresponds to averaging the difference between $S^{SC}_r$ and $-\text{Tr}[\rho^{SC}_r \log \rho^C_{m_r}]$ over runs of the experiment, where $\rho^{SC}_r$ is a `shadow’ that is constructed from a classically-simulated experiment in which one applies $U_r$ to $\rho^C_{m_r}$ and measures in the computational basis (to be contrasted with $\rho^S_r$, which is constructed from $U_r$ and $\rho_{m_r}$). The basic principle is that, while the above difference certainly converges to $\mathbbm{E}_m[S^{QC}_m]-\mathbbm{E}_m[S^{CC}_m]$, the errors in the two contributions are correlated and partially cancel one another.

It is important to note that in current quantum simulation platforms it is often simpler to make all measurements simultaneously and destructively \cite{monroe2021programmable,browaeys2020many,kjaergaard2020superconducting}. We must then ask which measurement-induced collective phenomena are observable given this constraint. Remarkably, even in such an experiment, our scheme allows one to probe the effect that measuring one subsystem has on the entanglement entropy of another subsystem, as in Refs.~\cite{bao2022finite,lin2022probing}. Moreover, given existing data from experiments involving randomized measurements \cite{brydges2019probing,joshi2020quantum,zhu2022cross}, through purely classical post-processing we can already study ensembles of post-measurement states. 

To see how this works, consider a tripartite system with subsystems denoted $A$, $B$, and $B'$, prepared in an entangled state, and suppose we are interested in how measurements of $B$ affect the entanglement entropy of $A$. In each run prior to measuring all degrees of freedom, it is necessary to act on $A$ with random unitary operations $U_r$ that will be used for shadows $\rho^S_r$. After measuring, the set of outcomes $m_r$ in $B$ should be used to construct a classical estimate $\rho^C_{m_r}$ for the density matrix of $A$ that is conditioned on outcomes in $B$ but not in $B'$, i.e. we simply do not use the information from the measurements in $B'$. By cross-correlating the shadow $\rho^S_r$ of $A$, which is obtained from the random unitary operations and measurement outcomes there, with $-\log \rho^C_{m_r}$, we arrive at the measurement-averaged entanglement entropy of $A$ as if $B'$ had not been measured. This scheme works because, if we do not use the outcomes of measurements in $B'$, then by averaging over runs of the experiment we convert these measurements into local dephasing events, and the effects of local quantum channels are strictly local. 

Our approach should be contrasted with probes of statistical properties of measurement records, most notably the cross-entropy benchmark studied in the context of random circuit sampling \cite{arute2019quantum}. The linear cross-entropy benchmark was also advocated a resolution of the post-selection problem for the measurement-induced entanglement transition in stabilizer circuits in Ref.~\cite{li2022cross}. The difference is that the linear cross-entropy benchmark is a cross-correlation between full probability distributions over measurement outcomes $m$, i.e. it is a cross-correlation of $P_m$ and $P^C_m$, where $P^C_m$ is the classical estimate for the probability to find the set of outcomes $m$, whereas the cross-correlations discussed in this work are between few-body quantities conditioned on measurement outcomes, e.g. $\rho_m$ and $\rho^C_m$. Focusing on few-body quantities has the advantage that in generic (non-stabilizer) systems their fluctuations are much smaller. For example, if $P^C_m$ is a product of probabilities for $M$ different measurement outcomes, there are situations in which the number of experimental runs required to converge the average of $P^C_m$ grows exponentially with $M$.

As an example of one of our quantum-classical cross-correlations, we have identified an observable signature of generic measurement-induced criticality. In particular, we showed that for a cross-correlation at the critical point the constraint of finite classical memory (parameterized by a finite bond dimension $\chi$) introduces a length scale proportional to $\chi^{1/\lambda}$ with $\lambda \lesssim 2$. The classical memory required for our scheme therefore increases only as $L^{1+2\lambda}$. This behavior could have been anticipated from the fact that the exact measurement-averaged half-chain entanglement entropy (calculated before one measures all bulk qubits) grows logarithmically with $L$, while the theoretical maximum half-chain entanglement of a MPS grows logarithmically with $\chi$.

This result hints towards a theory of finite entanglement scaling at measurement-induced criticality, following Refs.~\cite{tagliacozzo2008scaling,pollmann2009theory,pirvu2012matrix}. A theory closer in spirit to those works would describe how simulations, corresponding to `classical-classical' probes in our terminology, behave as $\chi$ is increased. This would answer the question of the classical memory required to study measurement-induced criticality on a classical computer. The behavior of $\mathbbm{E}_m[S^{QC}_m]$ that we have addressed instead describes the classical memory required to observe critical behavior in experiment. 

Our work highlights the measurement-induced entanglement transition as the threshold beyond which we cannot observe the effects of our measurements. This threshold is mirrored in the quantum advantage in random circuit sampling \cite{napp2022efficient} and emphasizes the fact that, although quantum simulators have an immense capacity to store information, the randomness in its extraction is a deep limitation \cite{aaronson2005quantum,bremner2011classical}. It is pertinent to ask whether, by cross-correlating quantum and classical simulations, we can learn more than from classical simulation alone. 
\\
\section*{Acknowledgements}
We are grateful to Sajant Anand, Yimu Bao, Matthew Fisher, Eun-Ah Kim, Yaodong Li, Katarzyna Macieszczak and Max McGinley for useful discussions, to Zack Weinstein for helpful comments as well as for collaboration on related work \cite{garratt2022measurements}, and to Matteo Ippoliti for pointing out the bound on the entanglement negativity discussed in Sec.~\ref{sec:negativity}. This research was supported by the Gordon and Betty Moore Foundation (S.J.G.), the U.S. Department of Energy, Office of Science, Office of High Energy Physics, under QuantISED Award DE-SC0019380 (S.J.G.), the National Science Foundation Quantum Leap Challenges Institute through grant number OMA-2016245 (E.A.) and a Simons Investigator award (E.A.). Numerical calculations were carried out using the Lawrencium computational cluster resource provided by the IT Division at the Lawrence Berkeley National Laboratory (Supported by the Director, Office of Science, Office of Basic Energy Sciences, of the U.S. Department of Energy under Contract No. DE-AC02-05CH11231).

%

\end{document}